\documentstyle[psfig,epsf]{article}
\newcommand{\dzero}{D\O\,\,}
\newcommand{\emis}{/\!\!\!\!E}

\begin{document}

\begin{center}
{\Large \bf Higgs and $B$ physics in Run II} \\

\vspace{4mm}

Valentin E. Kuznetsov\\ 
(for the \dzero Collaboration) \\ \ \\
{\it Department of Physics}\\
{\it University of California}\\
{\it Riverside, CA 92521-0413, USA}
\end{center}

\begin{abstract}
In Run II at the Tevatron, the major goal of the upgraded CDF and \dzero  
detectors is a Higgs search in the mass range of $110-200$ GeV.
They will also contribute significantly to $B$ physics. Among many
possibilities they will be able to measure rare decays of $B$ mesons and 
improve our knowledge of CP violation in $B$ system through study of 
$B$ mixing. Various aspects of Higgs and $B$ physics in Run II are 
discussed here.
\end{abstract}

%

\section{Introduction}

Both of the CDF and \dzero  detectors, are undergoing major upgrades in Run II, which
will start in March, 2001. The Tevatron will deliver 
approximately 2 fb$^{-1}$ integrated luminosity to each experiment, a factor
of 20 more than the Run I data samples. This will provide a unique opportunity 
for both experiments to search for the Higgs boson in the mass range of 110
to 200 GeV until new LHC data arrives.
At the same time, a wide range of other physics topics will be present.
Among these we expect to study various rare decays of $B$ mesons, 
search for CP violation and $B_{d,s}$ oscillations. 
In this note we discuss various aspects of Standard Model (SM)
Higgs production and $B$ physics in Run II.

\section{Experimental Layouts}

Details of the \dzero detector can be found elsewhere \cite{D0_detector}. 
Here, we briefly summarize the main features of the detector
relevant for Higgs and $B$ physics. 
The heart of the \dzero detector is a silicon tracking system (SMT), which consists of
six barrel segments with a disks in between and three more disks located
at each end of the tracker.
%
%
The barrel and disks are based on 50 $\mu m$ pitch silicon
microstrip detectors, providing spatial resolution $\sim 10\mu m$.
At each end of this system the two  large disks are placed in order to increase
$\eta_{det}$ coverage. The SMT system is enclosed in fiber tracker (CFT). 
They represent a complete robust tracking system of the detector.
\dzero detector will allow us to have momentum resolution at the level of
$\sigma(p_T)/p_T=0.02-0.05$ for low $p_T$ tracks 
with quite high tracking efficiency for charged particles at
$|\eta_{det}|<3$. Vertex reconstruction resolution is expected to be
$15-30\mu m$ in ($r-\phi$) plane for primary vertices and 
for secondary vertices it is expected to be $40\mu m$ in the ($r-\phi$) and
$100\mu m$ in the $(r-z)$ planes, respectively.
A major upgrade of the muon system together with central and forward
scintillators will allow us to trigger charged tracks.
Electron and muon identification will be possible in the central and forward regions. 

The CDF detector has similar capabilities such as a new silicon and central
outer trackers, plug calorimeter, muon chambers, and data acquisition
system. Owing an additional silicon layer near the beam pipe, a time of flight
system, they are expect to have slightly better vertex reconstruction
resolution \cite{CDF}.

\section{Higgs Production at the Tevatron}

\begin{figure}[thb]
\centerline{\psfig{figure=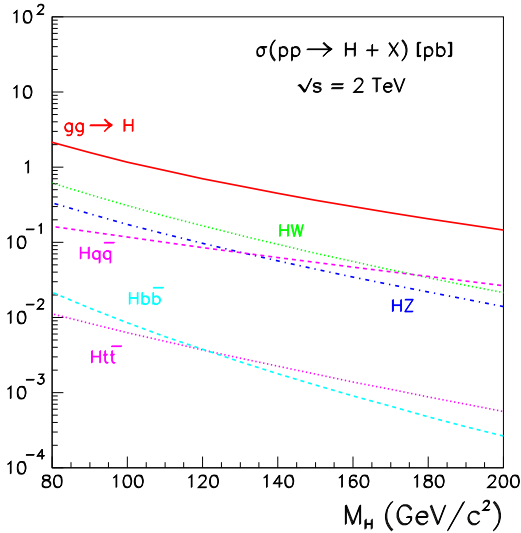,width=8.0cm}}
\caption{\label{Higgs_cs} The SM Higgs cross section for six production modes versus
Higgs mass.}
\end{figure}

\begin{figure}[thb]
\centerline{\psfig{figure=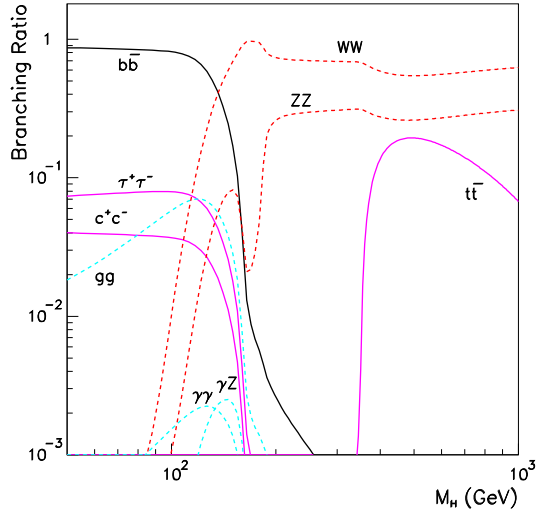,width=8.0cm}}
\caption{\label{Higgs_br} The branching fractions of various decay modes of the 
SM Higgs versus its mass.}
\end{figure}

In the Standard Model, Higgs bosons are expected to be produced in gluon fusion
or in conjunction with a $W$ or $Z$ boson. 
The expected cross sections at the Tevatron are shown in
Fig. \ref{Higgs_cs} \cite{Spira}.
Although the gluon fusion mode is expected to be largest contributor
to Higgs boson production, it will be  overwhelmed by the huge QCD background. 
Therefore pay most attention to the $WH$ and $ZH$ production modes.
The Higgs boson will mainly decay into $b\bar{b}$ and $WW$ final states for
the mass range below and above 135 GeV, as shown in Fig. \ref{Higgs_br} \cite{Spira}.

In the low mass range, $m_H\simeq 90-130$ GeV, in 90\% of the cases the
Higgs will decay into a $b\bar{b}$ pair giving a $(q\bar{q}, \ell^+\ell^-,
\ell\nu, \nu\bar{\nu}) b\bar{b}$ final state where the leptonic decay
of the $W$ or $Z$ will serve as a trigger. 
The high-$p_T$ lepton ($e$ or $\mu$) trigger is expected to use in selection 
of $HW$ mode where $W\rightarrow\ell\nu$. 
We assume that such a trigger will be efficient for
leptons with $p_T\geq20$ GeV. Also a large amount of missing transverse
energy $\emis_T\geq20$ GeV is expected. Two $b$-tagged central jets are
expected to survive in the ``loose'' and ``tight'' selection cuts.
The main background for
this mode will be $Wb\bar{b}$ and $WZ$ decays. For the 
$\nu\bar{\nu} b\bar{b}$ final state the selection criteria are based on
missing transverse energy from the escaping neutrinos and
two distinct $b$-jets. Here the main backgrounds come from $Zb\bar{b}$ and
$ZZ$ events. For the $\ell^+\ell^- b\bar{b}$ final state,
the trigger threshold for final leptons could be reduced to $p_T\geq10$ GeV.
The invariant mass of two outgoing leptons should be consistent with $Z$ boson mass. 
Two separate $b$-jets are also required in this case. The main background will 
come from real $Z$'s produced in conjunction with $b\bar{b}$ pairs, or from $W$'s
decaying hadronically. Finally, for the $q\bar{q}b\bar{b}$ case, the expected
background from QCD events is expected to be unreduceable. 
The cross sections of di-jet and four-jet events are expected to be of the order 
of ${\cal O}(10^6)$ and ${\cal O}(10^4)$ pb, respectively.

For the mass range, $m_H\simeq120-190$ GeV, where the Higgs boson produced
in conjunction with a vector bosons, it will mainly decay into
$W^*W^*$ states \footnote{Hereafter, the $W^*(Z^*)$ denotes a $W(Z)$ boson 
of either on- or off-mass-shell.}
with subsequent decay $(W,Z)W^*W^*\rightarrow \ell^\pm\nu\ell^\pm\nu jj$.
For this case selection criteria requires
two leptons with $p_T\geq10$ GeV having the same charge and
two separate jets with $p_T\geq15$ GeV and at least $10$ GeV $\emis_T$.
The main background in this case is $WZjj$ production.
Also it is possible to consider tri-lepton final states produced in
$W^\pm H\rightarrow\ell^\pm\nu W^* W^*\rightarrow\rightarrow
\ell\nu\ell\nu\ell\nu$ chain. The main attraction of this case is a small
background. One could find the best combination of trileptons and fit their 
invariant masses to be consistent with $W^\pm W^*W^*$ and $W^\pm W^* Z^*$ final states. 
Nevertheless we do not expect much to gain from this channel due to
its small branching fraction.

In the case of gluon fusion production of the Higgs boson the four possible combinations
are allowed:
\begin{eqnarray*} 
 H&\rightarrow& W^*W^* \rightarrow \ell\nu jj \quad{\rm and}\quad \ell^+\ell^-\nu\bar{\nu}, 
 \\ \nonumber
 H&\rightarrow& Z^*Z^* \rightarrow \ell^+\ell^- jj\quad{\rm and}\quad \ell^+\ell^-\nu\bar{\nu}.
\end{eqnarray*}
For the $\ell^+\ell^-\nu\bar{\nu}$ channel we expect to have a large
background from  $WW$, $WZ$, $ZZ$, and $t\bar{t}$ production. It can be
reduced requiring two leptons with $p_T\geq10$ GeV and $\eta_{det}$
constrain. One could possible use a transverse and a ``cluster mass'' 
mass constrains and perform likelihood analysis, see \cite{Fermilab_Higgs}. 
The rest of the two final states can be considered using proper cut for 
outgoing leptons and requiring two distinct jets.

\begin{figure}[htb]
\centerline{\psfig{figure=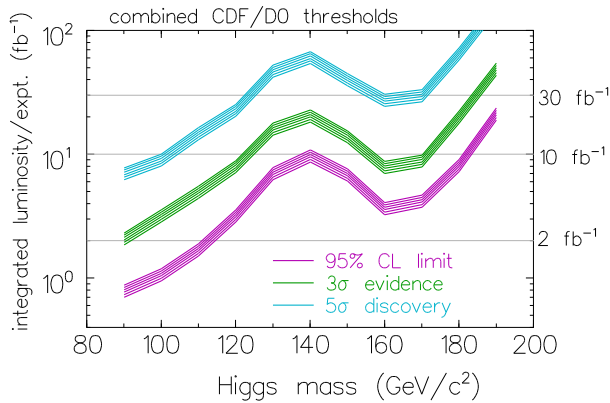,width=8.0cm}}
\caption{\label{Higgs_exp} Integrated luminosity delivered per experiment
required to either exclude at 95\% C. L. (bottom curve) or discover the SM Higgs at the 
$3\sigma$ (middle curve) or $5\sigma$ (top curve) level as a function of Higgs mass.
The theoretial uncertainties are already included in all curves.}
\end{figure}

Among various analyses underway in both collaborations,
the most promising ones are based on a neural networks technique. 
Various sets of kinematic variables is used 
to discriminate signal from background. 
Among them are transverse momentum of the isolated lepton, its pseudorapidity, 
the missing transverse energy of the event, 
the invariant mass of the final $b\bar{b}$-pair, and the separation 
between the $b$-tagged jets, the lepton and the first $b$-tagged jet, and
between the lepton and the second $b$-tagged jet. One of the important tasks is to find
the best possible $b$-tagging efficiency and mass resolution which are
expected to be at the level of 55-60\% and 10\%, respectively.
It looks like that it will be impossible to choose a golden channel, and
that success in a Higgs search will require a combination of all possible channels.
The details of various techniques as well as latest
results for Higgs search can be found in Ref. \cite{Fermilab_Higgs} and/or
under web \cite{WWW}.
As a result of preliminary analyses, we show in Fig. \ref{Higgs_exp}
the integrated luminosity delivered per experiment which would be
required to either exclude the SM Higgs at the level of 95\% or 
to make a discovery at the $3\sigma$ and $5\sigma$ levels.
The wide bands in the plot represent the calculated threshold plus 
uncertainties in the $b$-tagging efficiency, background rate, mass resolution, and
other effects. As the plot shows, in order to cover the full possible
spectrum of Higgs mass allowed at Tevatron energies, the total integrated luminosity
should be extended up to 30 fb$^{-1}$ per experiment. Even though a combination of all
channels as well as the data from both experiments are needed, new approaches and 
robust reconstruction algorithms will also be required.

\section{$B$ physics at the Tevatron}

The weak decays of $B_d$ and $B_s$ mesons play crucial roles
in the study of CP violation effects both within and beyond the Standard Model.
The CKM \cite{CKM} matrix elements, determined from 
various $B$ decay channels, can be represented in the Wolfenstein parameterization
\cite{Wolfenstein}  as a set of four parameters
$A, \lambda, \rho, \eta$. The parameters $A$ and $\lambda$
are known with good accuracy \cite{Mele}:
\[ \lambda=0.2196\pm0.0023,\quad |V_{cb}|=(39.5\pm1.7)\times10^{-3}\quad
   A=\frac{|V_{cb}|^2}{\lambda^2}=0.819\pm0.035.
\]
The $\rho$ and $\eta$ parameters can be extracted mostly from
four processes: CP violation in the neutral kaon system, oscillations
of $B^0_d$ and $B^0_s$ mesons, and charmless semileptonic $b$ decays.
The last three of these are the subject of great interest in the Run II $B$ physics
program.

\begin{figure}[thb]
\centerline{\psfig{figure=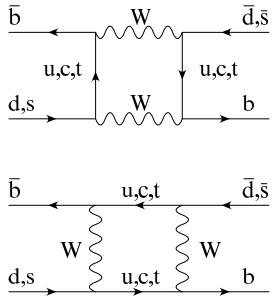,width=3.5cm}}
\caption{\label{BB} Feynman diagrams responsible for mixing in the $B$
systems.}
\end{figure}

The expected luminosity of the Tevatron, $2\times10^{32}$ cm$^{-2}$s$^{-1}$,
in Run II will lead to a huge rate of $b\bar{b}$ production, $\sim 10^{11}$
events/year.
This enormous statistics will allow us to study various $B$ decays modes, 
search for CP violation and $B_{d,s}$ mixing. 
Primary interest will focus on the study of CP-violation,
and related constraints on $|V_{td}/V_{ts}|$ from $B_s$ mixing.
Oscillations in $B$ system occurs because of high-order corrections, as
shown in Fig. \ref{BB}.
The light and mass eigenstates, $B_L$ and $B_H$, are
different from the CP eigenstates $B^0_q$ and $\overline{B^0_q}$:
\[ |B_L\rangle = p|B^0_q\rangle + q|\overline{B^0_q}\rangle \quad
   |B_H\rangle = p|B^0_q\rangle - q|\overline{B^0_q}\rangle,\quad
   q=d,s\;{\rm quarks}. \]
Unlike the kaon system, the mass difference
$\Delta m_q=m_{B_H}-m_{B_L}$ is the key feature of the physics.
Many analyses of $B^0_d-\overline{B^0_d}$ oscillations have been 
performed by several collaborations and their results have been combined to
give \cite{PDG}: $\Delta m_d = 0.472\pm0.017\; {\rm ps}^{-1}$.
%
%
For the case of $B^0_s$ mesons, due to their large mass difference $\Delta m_s$,
the $B^0_s$ oscillation frequency is thought to be much higher then the
well measured $B^0_d$ one. Existing data, mostly from CERN experiments,
exclude small values of the mixing parameter $x_s$,
$x_s=\Delta m_{B^0_s}/\Gamma_{B^0_s}>14.0$ at the 95\% CL \cite{PDG}.

Various decay modes of $B^0_s$ mesons are under investigation by the \dzero
collaboration. Among them
\begin{eqnarray*} 
  B^0_s &\rightarrow& D^-_s(K^-K^+\pi^-)\pi^+,\quad (B=1.1\times10^{-4}),
  \\ \nonumber
  B^0_s &\rightarrow& D^-_s(K^-K^+\pi^-)3\pi,\,\quad (B=2.8\times10^{-4}),
  \\ \nonumber
  B^0_s &\rightarrow& J/\Psi K^*,\;\;\;\quad\quad\quad\quad\quad (B=5.1\times10^{-6}),
  \\ \nonumber
  B^0_s &\rightarrow& D^-_s(K^-K^+\pi^-)\ell^+\nu,\;\;\,\, (B=1.1\times10^{-4}).
  \\ \nonumber
\end{eqnarray*}
The final data sample will be determine by the quality of the track and 
secondary vertex reconstruction algorithms. Final states of $B$'s
are expected to be fully reconstructed and tagged by the charge of the lepton
and reconstructed charm meson and/or kaon. The initial state required
for a $B_s$ mixing search can be tagged in one of two possible ways,
either by applying same-side tagging or opposite-side tagging techniques.
Using the good SMT resolution, we can tag $B$ decays using displaced secondary
vertices or tracks with large impact parameters.
For a 10\% $B$ reconstruction efficiency and with 12\% of $B_s$ 
having $p_T>0.5$ GeV and $|\eta_{det}|<1.5$ for all final state particles,
we expect approximately 2500 reconstructed events. We will be
sensitive for $x_s\leq20$. A new limit for
$|V_{ts}/V_{td}|$ can be establish using the well-known relation:
\[ \frac{\Delta m_s}{\Delta m_d}=\frac{m_{B_s^0}}{m_{B_d^0}}\xi^2
   \left|\frac{V_{ts}}{V_{td}}\right|^2
   =\frac{m_{B_s^0}}{m_{B_d^0}}\frac{\xi^2}{\lambda^2}
    \frac{1}{(1-\rho)^2+\eta^2}, 
\]
where theoretical uncertainties are included in the quantity $\xi$
\cite{Flynn,Mele}:
\[
   \xi=\frac{f_{B_d}\sqrt{B_{B_d}}}{f_{B_s}\sqrt{B_{B_s}}}=1.14\pm0.08.
\] 
The CDF detector is expected to obtain
a better constrain on the $x_s$ parameter. Using their time-of-flight
system, $K/\pi$ separation at the level of $2\sigma$ and opposite kaon
tagging, they expect to reach $x_s\sim60$ \cite{B_CDF}.

A second major component of the $B$ physics studies will be a CP violation
search in $B$ decays. It is well known that an asymmetry in
the $B$ system is generated if both decay amplitudes are nonzero and
if the weak decay phase, $\phi_{decay}$, is different from the mixing one,
$\phi_{mixing}$. Defining mass eigenstates as
$|B_1\rangle = p |B^0\rangle + q |\overline{B^0}\rangle$
and $|B_2\rangle = p |B^0\rangle - q |\overline{B^0}\rangle$
we have the following definition of $\phi_{mixing}$ and $\phi_{decay}$:
\[
  \frac{q}{p} = \sqrt{\frac{m_{12}^*-i\Gamma_{12}^*/2}
                           {m_{12}  +i\Gamma_{12}/2}}
              \simeq \frac{V_{tb}^* V_{td}}{V_{tb}V_{td}^*}
              =e^{-2i\phi_{mixing}},
  \quad
  \bar{\rho}(f)=\frac{\langle f|H|\overline{B^0}\rangle}
                     {\langle f|H|B^0\rangle}=\eta_f e^{-2i\phi_{decay}}
\]
Experimentally, we need to look for the asymmetry of the final state, $f$:
\[ A_{CP}(t) = \frac{ \Gamma\left(\overline{B^0_q}\rightarrow f\right)-
                      \Gamma\left(B^0_q\rightarrow f\right) }
                    { \Gamma\left(\overline{B^0_q}\rightarrow f\right)+
                      \Gamma\left(B^0_q\rightarrow f\right) }
             = \sin(2\beta)\sin(\Delta m_q t), \]
and try to measure the quantity $A_{obs}=D_{mix}D_{tag}D_{bgd}A_{CP}$.
Here the dilution factor $D_{tag}=1-2p_{misstag}$ is defined via the correct
tag probability $p_{misstag}$ and together with efficiency $\varepsilon$
defines the tag's effectiveness, $\varepsilon D^2_{tag}$; where the
mixing factor $D_{mix}=\sin(\Delta m_q t)=x_q/(1+x_q)$ and 
$D_{bgd}=\sqrt{S/(S+B)}$. The uncertainties on $\sin2\beta$:
\[ \sin2\beta={\rm Im}\left[
   -\left(\frac{V_{tb}^* V_{td}}{V_{tb} V_{td}^*}\right)
    \left(\frac{V_{cs}^* V_{cb}}{V_{cs} V_{cb}^*}\right)
    \left(\frac{V_{cd}^* V_{cs}}{V_{cd} V_{cs}^*}\right)
                        \right],
\]
are defined as
\[\delta(\sin 2\beta)=\frac{1}{D_{mix}D_{bgd}}\sqrt{1/(\varepsilon
                               D^2)_{tag}N_{rec}},\]
where $N_{rec}$ is the number of reconstructed events.
The three major blocks in $\sin2\beta$ comes from 
$B^0-\overline{B^0}$ mixing, 
$\left(\frac{V_{tb}^* V_{td}}{V_{tb} V_{td}^*}\right)$,
final decay fraction $\bar{\rho}(f)$, 
$\left(\frac{V_{cs}^* V_{cb}}{V_{cs} V_{cb}^*}\right)$ and
$K^0-\overline{K^0}$ mixing, 
$\left(\frac{V_{cd}^* V_{cs}}{V_{cd} V_{cs}^*}\right)$.
Flavor tagging efficiency will play a crucial role in final
purity of the samples. We have summarized it in Table \ref{FlavorTag}. 
 \begin{table}
 \begin{center}
 \begin{tabular}{rrrrr}
 \hline\hline
    & $\varepsilon D^2 (\%)$ & $\varepsilon D^2 (\%)$ & Relevant  & $\varepsilon D^2 (\%)$ \\
 Tag& measured               & expected               & \dzero    & \dzero \\
    & CDF Run I              & CDF Run II             & difference& capabilities \\ \hline
 same side   & $1.8\pm0.4\pm0.3$ & 2 & same & 2 \\
 soft lepton & $0.9\pm0.1\pm0.1$ & 1.7 & $\mu$, e ID coverage & 3.1 \\
 jet charge  & $0.8\pm0.1\pm0.1$ & 3 & forward tracking & 4.7 \\
 opp. side   &  & 2.4 & no $K$ id & none \\ 
 &&&&\\
 combined    & & 9.1 & & 9.8 \\
                                                           \hline\hline
 \end{tabular}
 \end{center}
 \caption{\label{FlavorTag} Flavor tagging efficiencies for 
 both CDF and \dzero detectors based on knowledge from Run I and
 MC studies \cite{Brun2}.}
 \end{table}
All numbers are based on our knowledge from Run I and MC studies. As
can be seen from the table our tag effectiveness will not exceed 10\%.
The golden mode is expected to be the decay of $B\rightarrow J/\Psi+K$ which is quite easy
to trigger when $J/\Psi\rightarrow \ell^+\ell^-$. The following
cuts can be applied in this case:
$p_T>1.5$ GeV for muon tracks, $p_T(K)>0.5$ GeV
and $|\eta_{det}|<2$. We expect to have approximately 40000
$B^\pm\rightarrow J/\Psi+K^\pm$ and 20000 $B^0\rightarrow J/\Psi+K^{0*}$ 
events with statistical errors dominating the systematics.
For the time-independent analysis, assuming 2 fb$^{-1}$ integrated luminosity,
$S/B\simeq0.75$, and a tag effectiveness $(\varepsilon D^2)_{tag}\simeq
9.8\%$, our expectation leads to $\delta(\sin2\beta)\simeq 0.04$ for
the case of $J/\Psi\rightarrow \mu^+\mu^-$ and
$\delta(\sin2\beta)\simeq 0.05$ for the $J/\Psi\rightarrow e^+e^-$ channel.
With 20 fb$^{-1}$ we will go down to $\delta(\sin2\beta)\simeq0.01$. Such
precision, together with other measurements, will tune the
position of the unitary triangle in the $\rho-\eta$ plane and help us better
understand the SM parameters.

Among other possibilities a large area of $B$ physics can be covered in
Run II: individual hadron masses and lifetimes ($B^\pm$, $B^0$, $B_s$, $\Lambda_b$)
search for $B_c$ meson ($B_c\rightarrow J/\Phi+\pi$)
search for rare b-decays ($b\rightarrow X_s\ell^+\ell^-$,
$b\rightarrow \ell^+\ell^-$, $B_s\rightarrow K^*\gamma$,
$B_s\rightarrow D_s^{*+}D_s^{*-}$), and a
non-SM CP violation search via the $B_s\rightarrow J/\Psi +\phi$
channel, but they are out of topic of this discussion.

\section{Conclusion}

As we have shown, the forthcoming Run II at the Tevatron collider is
expected to give us
a unique opportunity to study various exciting aspects of the SM, such as
Higgs search and CP violation. Thanks to the huge effort from many people,
the \dzero and CDF detectors are on the final road and ready to start
collecting data in March 2001. The main improvements in Run II that
make us optimistic for the physics are based on increased integrated
luminosity, at least factor of 20x more then in Run I, and major upgrades 
of both detectors. For both goals, Higgs search and $B$ physics, the main
gains will come from $b$ identification and from the $b\bar{b}$ 
invariant mass measurement.
Now our focus is concentrated on building a robust, reliable reconstruction
algorithm, in order to fully explore the detector capabilities.
Preliminary studies by the Higgs search working group show that there is
no golden channel in the analysis, and in order to be successful,
we need to combine all data and all decay modes from both experiments.
Initial luminosity of the Tevatron, 2 fb$^{-1}$, will allow us to
explore Higgs mass at the level of LEP2, but with 
10 fb$^{-1}$, we will be able to exclude at 95\% C.L. an SM Higgs
up to $\sim 180$ GeV and with 20 fb$^{-1}$ we could see its evidence
at the level of 3$\sigma$. \dzero and CDF will significantly contribute to
$B$ physics in Run II. The main focus of our attention will be the $B_s$ oscillation
search and CP violation. Several different modes are under investigation.
We expect to improve the lower limit on the $x_s$ parameter up to a level of
20 (D\O) and 60 (CDF) and extend our knowledge about the unitary triangle
by measuring $\sin2\beta$ with an accuracy of 0.01-0.05.

\section{Acknowledgments}

This talk would not have been possible without the help of a large number of
people. I would especially thanks the wise comments of my
colleagues Ella Barberis, Anna Goussiou, Maria Rocco, Rick Jesik, 
Su Yong Choi, John Ellison,  Ann Heinson and Dennis Shpakov. 
Finally, I would like to thanks all members
of organizer committee, especially Natalia Sotnikova and Edward Boos, 
for choosing a nice place for the meeting and for their help during
the conference.


\end{document}